# A General-Purpose Neuromorphic Sensor based on Spiketrum Algorithm: Hardware Details and Real-life Applications

MHD Anas Alsakkal, Runze Wang, Piotr Dudek, and Jayawan Wijekoon, *Member, IEEE*

***Abstract*—Spike encoding can transform sensory signals into sparse, structured event streams, reducing the data handled by downstream processors. However, encoding is often performed in software, limiting efficiency near the sensor. Spiketrum is a bio-inspired, general-purpose encoder that applies matching pursuit over a configurable kernel set. Through iterative residual feedback, it extracts salient features and represents each by its kernel identity, temporal position, and intensity. The resulting codes support signal reconstruction and can be processed by spiking or conventional models. An adjustable spike rate allows representation detail to be matched to application and hardware budgets. This paper presents a scalable, area-optimized hardware implementation that performs time-domain correlation using an 18-bit fixed-point datapath and 1–40 parallel multiply-accumulate lanes. On an FPGA, the encoder provides 5–212 spikes/s, with identical codes across lane configurations. Compared with the frequency-domain performance-optimized design, the eight-lane implementation uses 88% fewer BRAMs, 95% fewer DSP slices, and 71% fewer LUTs, with 71% lower power. A single-lane core was physically designed in TSMC 180-nm technology, with a standard-cell area of 0.062 mm² and total power of 63 mW at 200 MHz. Sound and ECG classification with spiking and non-spiking models demonstrates applicability across signal domains and processing paradigms. These results support Spiketrum's potential as an efficient sensory front end for edge systems, reducing data movement and downstream processing. Future work will assess broader modalities, including touch and olfactory sensing.**

***Index Terms*— ECG classification, FPGA implementation, integrated circuit, neuromorphic sensing, sound classification, spike encoding.**

## I. Introduction

BRAIN-INSPIRED computation has paved the way for significant advancements through Spiking Neural Networks (SNNs). SNNs have attracted considerable attention due to their biologically inspired architectures and energy-efficient computational paradigms [1]. Unlike conventional neural networks, SNNs process information using discrete spike events. These spikes do not encode information through explicit amplitude values; instead, information is conveyed through their temporal occurrence and spatio-temporal relationships. However, the conversion of external data into spike-based representations—commonly referred to as spike encoding—has largely been implemented in software or through hybrid software–hardware approaches, which can limit the efficiency gains achievable with fully dedicated hardware implementations [2]. Although substantial progress has been made in the development of spike-based processing hardware over the past decade, many encoding strategies continue to rely on conventional processing techniques, constraining the overall system efficiency [2].

While specialized spike encoding hardware exists, its rigid implementation often confines it to a single encoding scheme, lacking the adaptability required for dynamic reconfiguration [3-5]. Bohte et al. utilize temporal population coding to chronologically encode input data [6], whereas temporal encoding has also been applied for the classification of the MNIST dataset consisting of handwritten digit images [7]. On the other hand, rate encoding has been harnessed by Liu et al., Querlioz et al., and Bagheri et al. for the recognition of images containing digits [8-10]. The efficacy of SNNs in real-world applications is intricately tied to the chosen spike encoding scheme [11]. Therefore, this limitation underscores the need for a more versatile and adaptable spike encoding approach.

Designing neuromorphic architectures that reproduce detailed functions of complex biological sensory systems could provide insight into sensory processing. However, implementing such architectures in hardware may not be efficient in power and area. Therefore, introducing a higher level of abstraction in the model, while maintaining the ability to design and implement the model in the latest low-power electronic technologies, could enable a more efficient, compact and stable sensor for computing machines and neural implantable devices. The algorithm of Spiketrum [12] is designed to characterize and convert time-varying analog signals into highly efficient spatiotemporal spike patterns. It offers a sparse and efficient coding scheme, allowing precise control over spike rates.

In earlier research [13], we introduced a "performance-optimized" methodology for implementing Spiketrum, which emphasizes processing speed by harnessing additional resources to alleviate computational burdens. This approach enables real-time processing on a field-programmable gate array (FPGA), offering improved management of computational tasks and enhanced performance, particularly in scenarios that benefit from high-fidelity input representation and increased spike rates. Nevertheless, this implementation

Manuscript received Feb XX 2026; revised Mar XX, 2026; accepted Apr XX, 2026. Date of publication May XX, 2026; date of current version Jun XX, 2026.
The authors are with the Department of Electrical and Electronic Engineering, The University of Manchester, Manchester, UK. Corresponding author *Jayawan Wijekoon* (e-mail: jayawan.wijekoon@manchester.ac.uk). Color versions of one or more of the figures in this article are available online at http://ieeexplore.ieee.org.

may not be appropriate for applications where factors such as power consumption and hardware footprint are paramount, as seen in cochlear implants, for example.

Thus, in this work, we present an "area-optimized" hardware implementation of Spiketrum, focused on conserving hardware resources and minimizing the overall implementation footprint. Relative to the frequency-domain performance-optimized implementation [13], this area-optimized design in its eight-lane reference configuration (Section II-A) uses 88% fewer block RAMs (BRAMs), 95% fewer digital signal processing (DSP) slices, and 71% fewer look-up tables (LUTs), with 71% lower estimated power.

The main contributions are: (i) a time-domain area-optimized Spiketrum architecture scalable from 1 to 40 parallel multiply-accumulate (MACC) lanes, with lane count controlling throughput while preserving identical codes; (ii) an FPGA comparison with the performance-optimized Spiketrum implementation and other hardware spike encoders; (iii) physical design of a single-lane core in TSMC 180-nm technology; and (iv) sound and electrocardiogram (ECG) classification using spiking and non-spiking classifiers.

The remainder of the paper is organized as follows. Section II presents the Spiketrum algorithm and FPGA implementation; Section III describes the IC design; Section IV details the evaluation methodology; Section V presents the classification results; Section VI compares the performance- and area-optimized implementations with recent hardware encoders; and Section VII concludes the paper.

## II. Spiketrum: Design Considerations and Implementation Details

Spiketrum [12] encodes a time-varying signal into spikes by matching pursuit [14] over a dictionary of 40 Gammatone kernels $\varphi_m$, each 1,353 samples long. The signal is processed in segments of 696 samples (43.5 ms at 16 kHz). In each iteration i, the segment $x(t)$ is convolved with all kernels, and the best match defines a code $(m_i, \tau_i, s_i)$: the kernel index, the time shift and the convolution intensity. The scaled and shifted kernel component $s_i \varphi_{m_i}(t-\tau_i)$ is then subtracted from the segment, and the residual $x_{new}(t)$ is encoded in the next iteration, so that the dominant features are encoded first. After k iterations, the segment is described by the code set:

$$S = [(m_1, \tau_1, s_1), \ldots, (m_k, \tau_k, s_k)] \quad (1)$$

Each code describes a small waveform with the shape of kernel $m_i$, located at $\tau_i$ and scaled by $s_i$. Superposing these waveforms reconstructs the segment, and the reconstruction error $\epsilon(t)$ decreases as k increases; it also depends on the kernel set. Intensity-to-Place coding finally maps every code to a spike on one of 120 output channels. Since each code produces one spike, k sets the spike rate, which is expressed in spikes per segment (sps). The hardware implements these steps in three stages, Feature Extraction, Residual Computing, and Intensity-to-Place coding, as shown in Fig. 1.

Feature Extraction finds the best-matching kernel and generates the code, Residual Computing removes the encoded component from the segment, and Intensity-to-Place Coding converts the code into a spike. Fig. 2 shows the hardware output for a cymbal tone from the MedleyDB dataset [15], sampled at 16 kHz and encoded at 1000 sps: the two input segments (Fig. 2(a)), the generated codes, plotted as kernel index against time shift with the marker size indicating the intensity (Fig. 2(b)), and the resulting spikes on the 120 output channels (Fig. 2(c)).

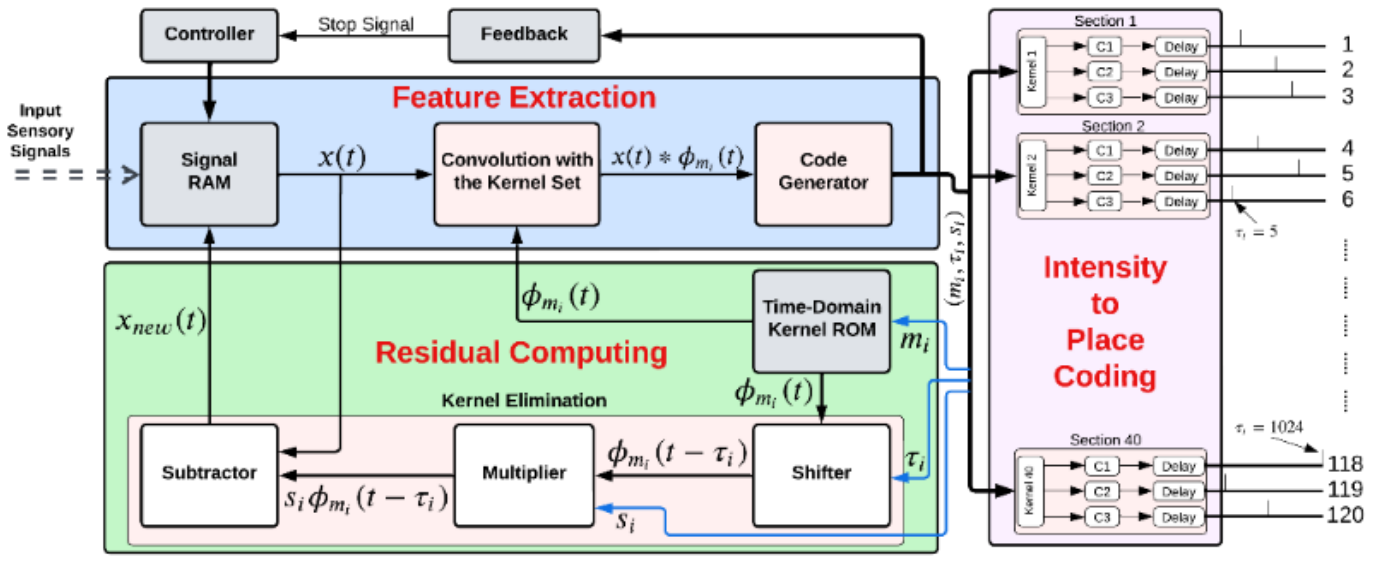

**Fig. 1.** Hardware architecture of the Spiketrum encoder. Feature Extraction and Residual Computing generate codes, which Intensity-to-Place Coding maps to spikes.

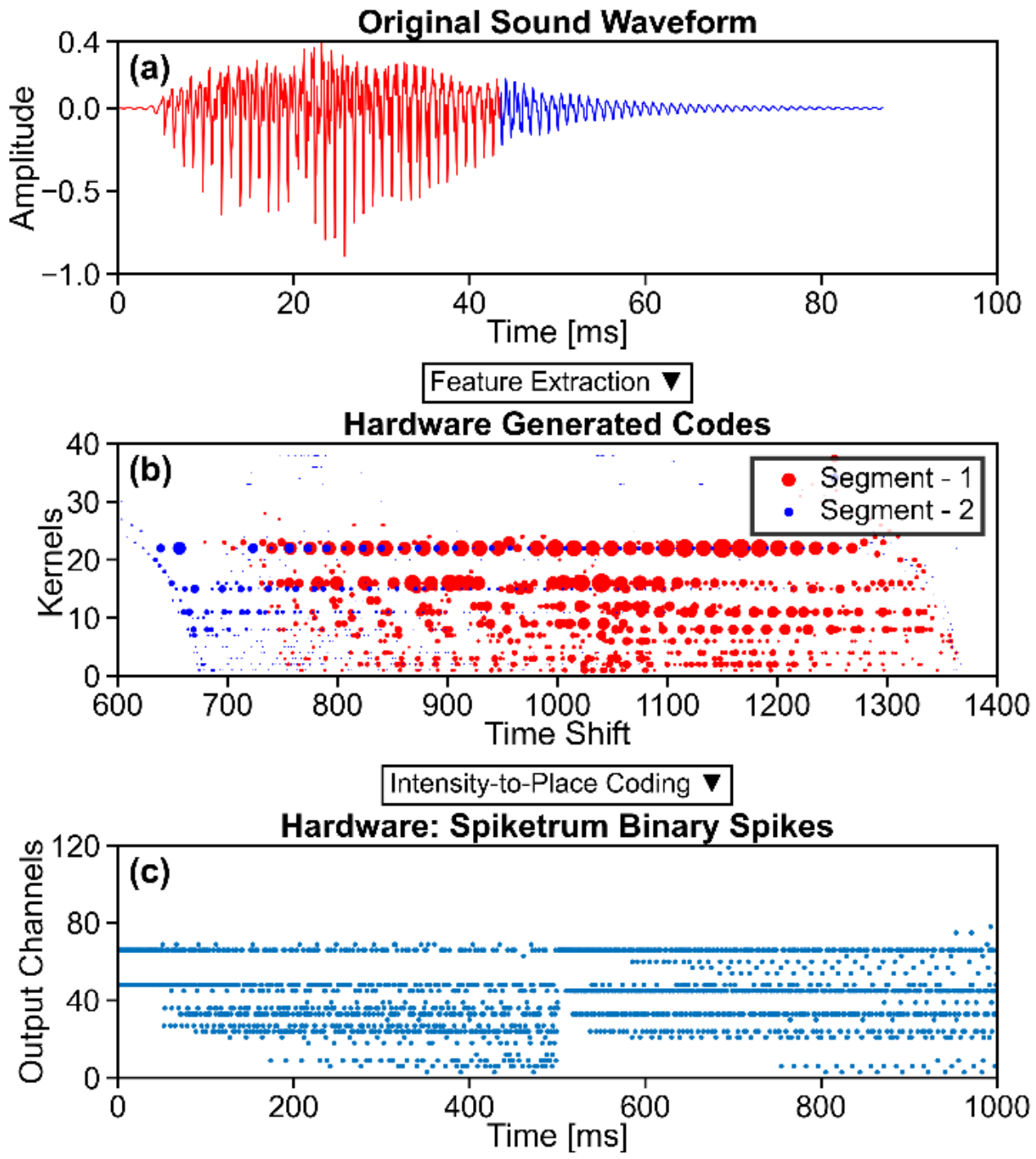

**Fig. 2.** Hardware output for a two-segment audio signal at 1000 sps. **(a)** Input waveform. **(b)** Generated codes (kernel index versus time shift; marker size indicates intensity). **(c)** Binary spikes across the 120 output channels.

The encoder is implemented on an Opal Kelly XEM7310 board with a Xilinx Artix-7 FPGA (Fig. 3). The host streams the input segments to the FPGA over USB 3 and reads back the codes and the output channel of every spike over the same interface, so that they can be processed by a spiking or a conventional (non-spiking) classifier. The board also carries microphone circuits for live audio input and a USB-C connector that facilitates direct access to the 120 spike channels over the Address-Event Representation (AER) protocol.

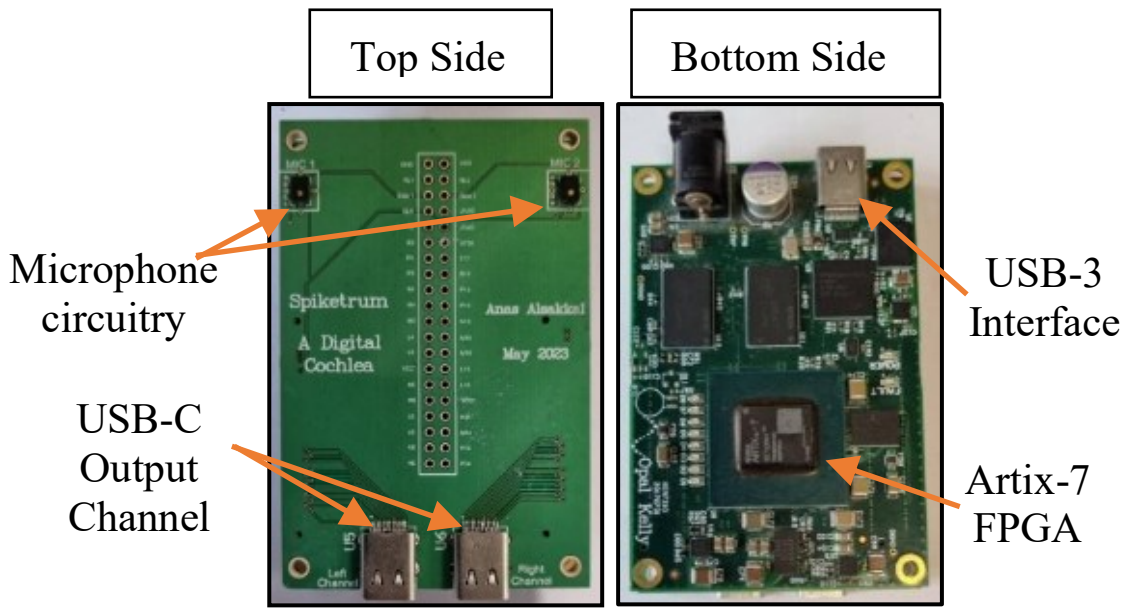

**Fig. 3.** Real-time FPGA prototype, shown from the top and bottom. Audio is supplied through a microphone or USB 3, and 120 spike channels are available via USB-C.

The convolution of the segment with all kernels dominates the computation and can be implemented in two ways (Fig. 4). The performance-optimized design [13] computes it in the frequency domain: the fast Fourier transform (FFT) of the segment is multiplied by the stored FFTs of the kernels and transformed back by an inverse FFT (Fig. 4(b)). At 200 MHz, it generates up to 80 spikes per segment (1,840 spikes/s) in real time, which suits applications that need high spike rates, such as autonomous vehicles or brain–computer interfaces (BCIs). In [16], it was benchmarked against three state-of-the-art spike encoders (MAP, Spectrogram and Lauscher) and reached competitive accuracy with substantially fewer resources.

The area-optimized design proposed here computes the convolution directly in the time domain with L parallel multiply-accumulate (MACC) lanes (L = 1 to 40), each built around one DSP slice. It needs neither FFT cores nor a frequency-domain kernel memory, and the number of lanes trades spike rate against resources: with eight lanes, a spike takes 23.6 ms, i.e., about 1.8 spikes per segment. The area-optimized design therefore targets applications in which footprint and power matter more than spike rate, such as cochlear-implant front ends; higher rates require more lanes or the performance-optimized design.

Both designs use fixed-point arithmetic, which needs less power and area than floating point [17]. The performance-optimized design uses 34 bits because the FFT needs a large dynamic range [13]. The area-optimized design uses 18 bits: the signal and kernel samples are 1.17 numbers (one sign bit and 17 fractional bits), and the convolution sums, intensities and scaled kernels are saturated 2.16 numbers. The 18-bit word matches the 18 × 18 multiplier of the DSP48E1 slices [18] and keeps the registers, memories and comparators small. All memories are generated with the Xilinx Block Memory Generator in Native mode, which uses fewer FPGA slices than an AXI4 interface. The remainder of this section describes the three stages of the area-optimized implementation.

### *A. Feature Extraction*

Each segment is loaded into the Signal RAM (Fig. 1), a simple dual-port RAM of 696 × 18 bits (12.5 kb) that holds the segment and, after every iteration, its residual. The 40 kernels are stored in the time-domain (T-D) Kernel ROM, a single-port ROM of 974 kb (about 122 kB). Small finite-state machines (FSMs), shown as the Controller in Fig. 1, generate the memory addresses of the convolution and of the residual update and repeat the iterations of the segment.

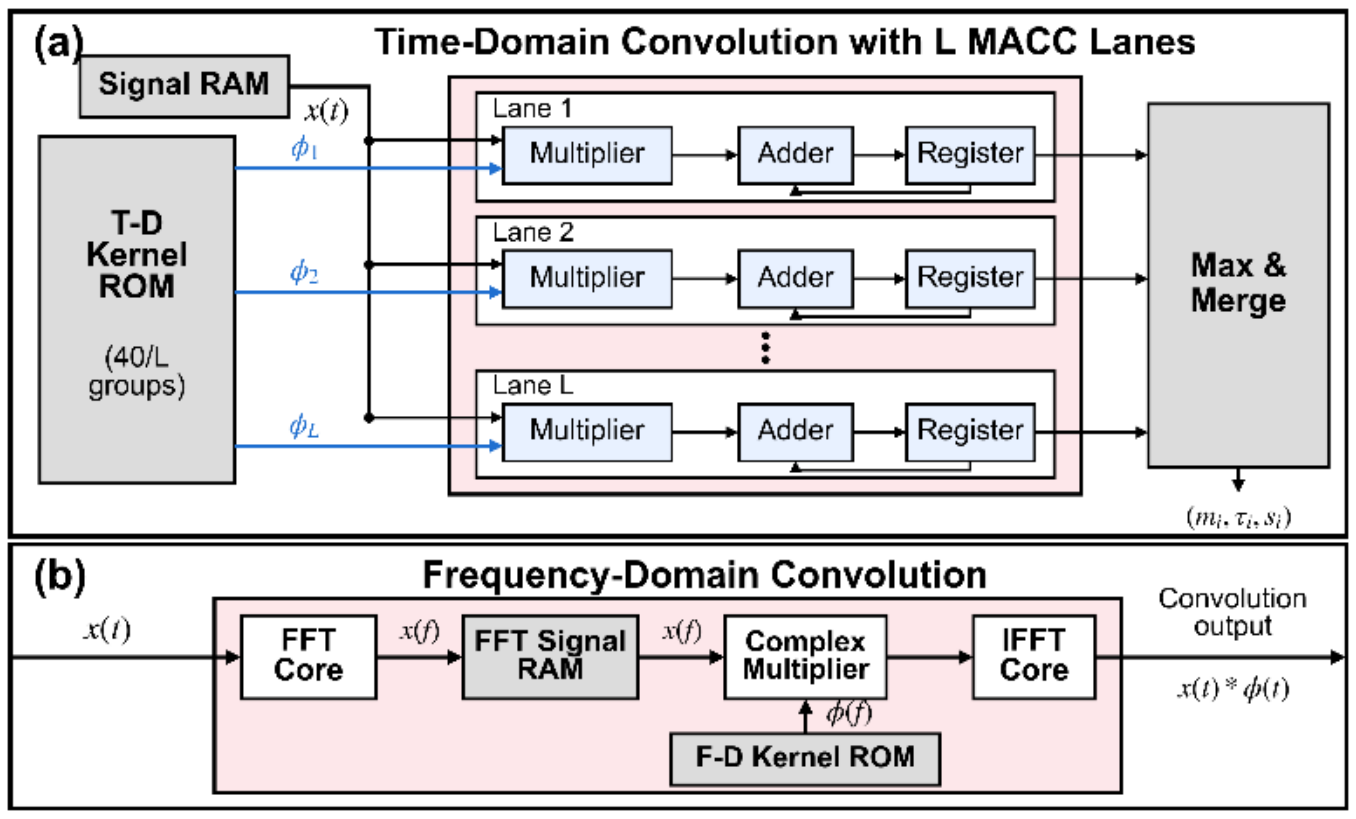

**Fig. 4.** Block diagrams comparing the proposed time-domain and frequency-domain implementations of the convolution operation. **(a)** Time-domain convolution uses L parallel MACC lanes (L = 1 to 40). Each lane multiplies the shared input by one kernel from the grouped ROM and accumulates the products; the largest lane result forms the code. **(b)** Frequency-domain implementation: the FFT core transforms the input signal $x(t)$ into $x(f)$, which is stored in the FFT Signal RAM; the complex multiplier multiplies $x(f)$ by the stored FFT of the kernel set $\varphi(f)$, and the inverse FFT (IFFT) core returns the convolution output. Further details are given in [13].

The convolution is computed by the L MACC lanes of Fig. 4(a), where L divides the number of kernels (L = 1, 2, 4, 5, 8, 10, 20 or 40). The T-D Kernel ROM is organized in 40/L groups, and each ROM word holds the same sample of the L kernels of a group, so that one read supplies all lanes. During a sweep, the segment slides along the kernels of one group: in every clock cycle, all lanes receive the same signal sample and the sample of their own kernel, and each lane accumulates the products into the convolution sums of its kernel at every time shift. A sweep takes 944,363 clock cycles (696 × 1,353 MACC steps plus pipeline overhead), and a code requires 40/L sweeps, so the spike rate grows linearly with L while the Signal RAM bandwidth stays the same. Each lane is a five-stage pipeline (input register, DSP multiplier, rounding, saturation and saturating accumulation) that runs at 200 MHz.

Table I lists the resources, power and speed for different numbers of lanes. Each additional lane adds one DSP slice and about 115 LUTs, whereas the block RAM usage stays between 32 and 35.5 tiles because only the aspect ratio of the kernel ROM changes. Power and resource utilization were estimated with Xilinx Vivado [19], as direct measurements of the FPGA logic are impractical.

The eight-lane configuration, used as the FPGA reference in the rest of the paper, consumes an estimated power of 401 mW at 200 MHz (133 mW static and 267 mW dynamic). The 40 lanes based design can provide up to 9.2 sps (212 spikes/s) so can support applications where real-time high spike rate is required. The single-lane core was taken through the IC design flow (Section III).

TABLE I
AREA-OPTIMIZED DESIGN WITH L MACC LANES

| Lanes | LUTs | DSP slices | Power (mW) | Time per spike (ms) | Energy per spike (mJ) | Spikes per segment |
|---|---|---|---|---|---|---|
| 1 | 1,801 | 4 | 315 | 189 | 60 | 0.2 |
| 5 | 2,316 | 8 | 363 | 37.8 | 13.7 | 1.2 |
| 8 | 2,551 | 11 | 401 | 23.6 | 9.5 | 1.8 |
| 10 | 2,819 | 13 | 417 | 18.9 | 7.9 | 2.3 |
| 20 | 3,979 | 23 | 482 | 9.4 | 4.6 | 4.6 |
| 40 | 6,339 | 43 | 607 | 4.7 | 2.9 | 9.2 |

Vivado 2018.3 estimates for an Artix-7 FPGA at 200 MHz. Energy per spike equals power times time per spike; spikes per 43.5-ms segment equal 43.5 ms divided by time per spike.

The Code Generator keeps, for every lane, the largest sum found so far together with its kernel index and time shift, using one 18-bit comparator per lane. After all kernels have been processed, it selects the largest of the lane maxima as the code. When two values are equal, the lower kernel index, and then the earlier time shift, is chosen, so the codes do not depend on the number of lanes L. The code ($m_i$, $\tau_i$, $s_i$) is then passed to Residual Computing and Intensity-to-Place Coding.

After every code, the Feedback block (Fig. 1) decides whether the encoding of the segment continues. In this design, the decision is made by a single comparator. If the intensity of the code is below a threshold set by the host (0.00125 by default), no further iteration is started after its residual update; otherwise, encoding continues up to a predefined number k of codes per segment. Segments that contain no significant signal, such as pauses between words, are thus represented by a single spike, and no time or power is spent on insignificant features. The host can also disable this halting to obtain a fixed number of spikes per segment. Since it acts on each code during encoding, the comparator can be replaced by more advanced mechanisms. One example is top-down control inspired by the descending (efferent) pathways from the auditory cortex to the cochlea, which adapt the sensitivity of the sensory periphery to the task and the environment.

### *B. Residual Computing*

Residual Computing removes the component of the latest code from the segment, $x_{new}(t) = x(t) - s_i\varphi_{m_i}(t - \tau_i)$ for t = 0, …, 695. In the area-optimized design, the shift requires no hardware of its own: it is realized by the addressing of the kernel ROM, and one residual-update unit performs the multiplication and the subtraction.

For every sample t, the residual address generator reads $x(t)$ from the Signal RAM and the sample at address $\tau_i - 1 - t$ of kernel $m_i$ from the T-D Kernel ROM; addresses outside the kernel (0 to 1,352) return zero. This realizes positive and negative shifts, including the truncation of the kernel at the segment boundaries, in a single pass of 696 sample pairs and without a shift memory.

The residual-update unit selects the sample of kernel $m_i$ from the L samples of the ROM word, multiplies it by $s_i$ in one DSP48E1 slice (18 × 18 two's-complement multiplication [18]), rounds and saturates the product to the 2.16 format and subtracts it from $x(t)$ in the logic fabric. These operations form a seven-stage pipeline (two stages for the lane selection and one each for the input registers, product, rounding, saturation and subtraction) that runs at 200 MHz. The residual is written back to the Signal RAM in the 1.17 format and replaces the segment in the next iteration. One update takes about 700 clock cycles, which is negligible compared with the convolution sweeps.

### *C. Intensity-to-Place Coding*

Intensity-to-Place Coding assigns N output channels, or fibers, to every kernel, each representing an intensity level. Statistical analysis [12] showed that N = 3 gives high sparsity at a practical hardware cost, with the logarithmically spaced levels $C_1$ = 0.0065, $C_2$ = 0.4115 and $C_3$ = 25.8744; the 40 kernels therefore give 120 output channels.

The Spike Generator subtracts the three levels from $s_i$ with three subtractors and selects the level $l$ closest to $s_i$. The spike is placed on output channel $3m_i + l$, where $m_i = 0, \dots, 39$ and $l = 1, 2, 3$, so that its place encodes both the kernel and the intensity, while its time is given by $\tau_i$. A delay counter emits the spike on its channel after a delay equal to $\tau_i$. Spikes are also communicated to the host over USB 3.

## III. SPIKETRUM: IC IMPLEMENTATION

The replication of a general-purpose neuromorphic sensor using silicon technology is driven by its diverse applications in domains such as BCIs, cochlear implants, and portable devices requiring low-power signal processing. These capabilities are particularly crucial for tasks like speech recognition on edge devices. The FPGA prototype implements the time-domain Spiketrum encoder and supports configurable multiply-accumulate (MACC) lane counts. Its programmability enables rapid prototyping and evaluation of throughput–resource trade-offs. To assess the architecture's feasibility in custom silicon, a single-lane core was also taken through the physical-design flow in TSMC 180-nm CMOS. This IC layout provides a proof of concept; multi-lane configurations are evaluated on the FPGA.

### *A. Chip Architecture*

The IC core comprises the MACC, code-generation, spike-generation, and residual-feedback units. It uses an 18-bit fixed-point datapath, a single clock domain, and synchronous reset. Signal RAM and the time-domain kernel ROM are off-chip; signal and kernel samples are streamed into the core, and selected pipeline signals are exposed at output pads for characterization.

The netlist contains 4,203 cells at 200 MHz, 3,283 at 100 MHz, and 3,287 at 50 MHz. The 200-MHz netlist includes 392 flip-flops and 14 clock-gating cells. One correlation sweep requires 944,363 cycles. Since a single-lane code requires 40 sweeps, code generation takes 37.8 million cycles, or 189 ms at 200 MHz. RTL simulations match the double-precision reference in kernel selection and time shift, with amplitude differing by no more than two least-significant bits.

### *B. Physical Design Flow*

The RTL was synthesized using Cadence Genus and TSMC 180-nm libraries. Synthesis enabled clock gating and targeted a 200-MHz clock, with a 5-ns period and 0.2-ns uncertainty. Cadence Innovus was used for floorplanning, placement, clock-tree synthesis, routing, antenna-diode insertion, and post-route setup/hold optimization.

The floorplan uses a 196-pad frame, including 184 signal pads at 90-µm pitch. Power rings and stripes were added, and clock-tree synthesis used a 150-ps skew target across 392 clock sinks. The core logic uses the tcb018gbwp7t standard-cell library; tpd018bcdnv5 provides I/O cells, and tpv018v provides bond pads.

### C. Implementation Results

The digital core targets input signals in the 20-Hz to 8-kHz band. Its 18-bit datapath corresponds to a nominal digital dynamic range of approximately 108 dB. Fig. 5 shows the IC layout, and Table II summarizes cell count, area, post-layout power, and energy per spike at 50, 100, and 200 MHz.

At 200 MHz, the estimated total power is 63.16 mW, and the 189-ms time per spike corresponds to 11.9 mJ per spike. At 100 and 50 MHz, estimated power falls to 26.05 and 13.16 mW, while time per spike increases to 378 and 756 ms; energy per spike is 9.8 and 9.9 mJ, respectively. Thus, reducing the clock rate lowers power but increases latency, while energy per spike remains within 9.8–11.9 mJ. A block-level power breakdown attributes approximately half of clocked-logic power to the MACC unit, with the multiplier and subtractor contributing about 10% each.

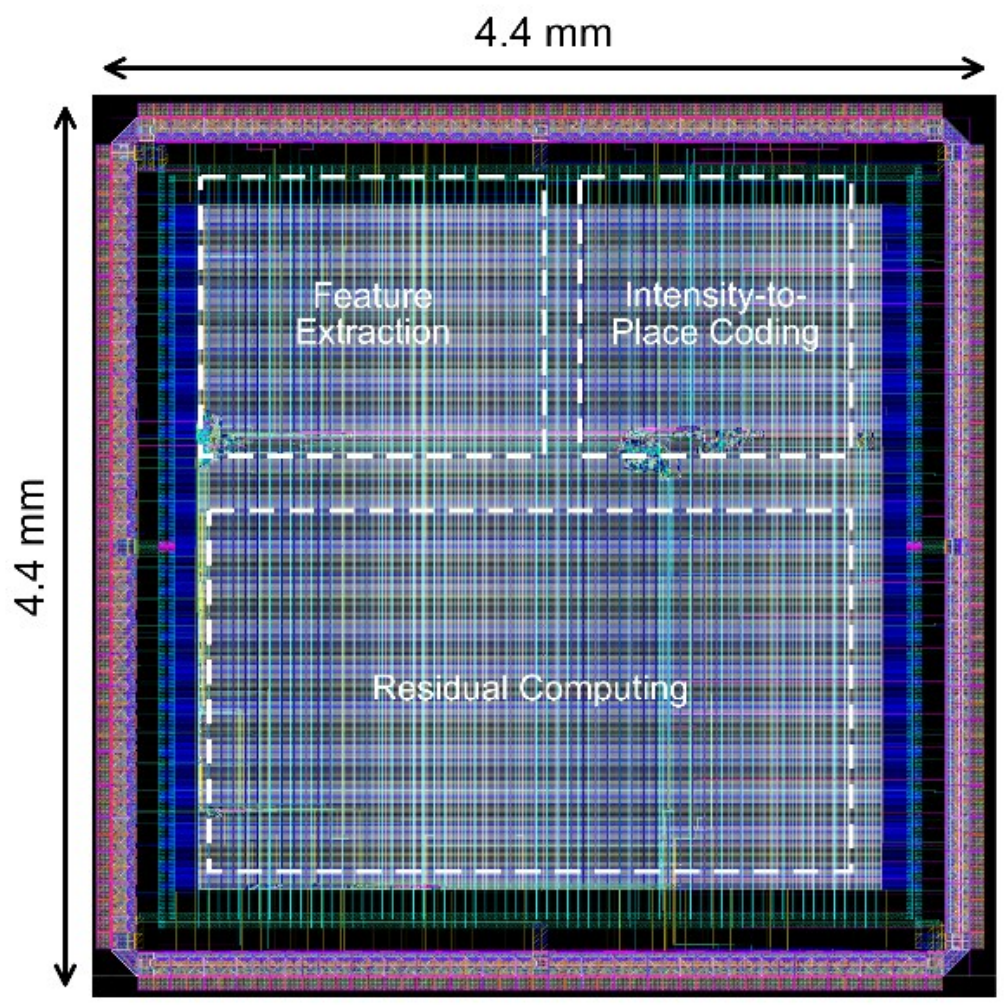

**Fig. 5.** An image of the layout of the proposed Spiketrum implementation.

Table II summarizes cell count, area, power, and energy per spike at 50, 100, and 200 MHz. At 200 MHz the IC consumes 63.16 mW, which corresponds to 0.32 nJ per multiply-accumulate step and 11.9 mJ per spike (189 ms per spike). At lower clock rates, power falls while spike-generation time increases, leaving energy per spike approximately constant. The MACC unit accounts for about half of clocked-logic power; the multiplier and subtractor each account for about 10%. Because the signal and kernel memories are off-chip, these results characterize the encoder core rather than a complete sensor system.

TABLE II
CHARACTERISTICS OF THE IC OF SPIKETRUM FOR VARIOUS OPERATING FREQUENCIES

| Clock Speed | | 200 MHz | 100 MHz | 50 MHz |
|---|---|---|---|---|
| Cell Count (cell) | | 4203 | 3283 | 3287 |
| Area (µm$^2$) | | 3,162 | 3,151 | 3,152 |
| Power Consumption (mW) | Internal | 42.83 | 18.34 | 9.28 |
| | Switching | 20.32 | 7.71 | 3.88 |
| | Leakage | 0.0006 | 0.0005 | 0.0005 |
| | Total | 63.16 | 26.05 | 13.16 |
| Performance | Time per spike (ms) | 189 | 378 | 756 |
| | Energy per spike (mJ) | 11.9 | 9.8 | 9.9 |

## IV. METHODOLOGY

In this section, we describe the methodologies used to evaluate the proposed area-optimized FPGA-based Spiketrum design across two real-world applications: sound classification and ECG classification. The evaluation focuses on comparing spiking and non-spiking neural network classifiers in terms of classification accuracy, computational efficiency, and training complexity.

For ECG classification, Spiketrum is used as a front-end spike encoder, followed by a small resource-efficient Multilayer Perceptron (MLP) classifier. The MLP is designed in conjunction with a Temporal Averaging preprocessing technique, which aggregates spike activity over time to produce compact feature representations suitable for low-complexity classification. This combination exploits the temporal characteristics of ECG signals to reduce computational overhead while maintaining competitive classification performance. In addition to the proposed MLP-based approach, the evaluation includes conventional deep learning models such as Convolutional Neural Networks (CNNs) and Long Short-Term Memory (LSTM) networks, as well as neuromorphic models including Recurrent Spiking Neural Networks (RSNNs) and Adaptive Recurrent Spiking Neural Networks (aRSNNs), enabling a comprehensive performance comparison.

The classification and evaluation results are presented in the next section (Section V), highlighting the efficacy of each model and preprocessing technique across the two application domains.

### A. Preprocessing Techniques

Two preprocessing techniques were employed on the spike sequences:

**Raw Spike Input Method:** Directly feeds unprocessed spike sequences to classifiers.

**Temporal Averaging Method:** Computes the average spike activity over time per channel, resulting in a static 120-dimensional vector. For ECG signals, the averaging is performed over the entire input window, corresponding to a 2-second segment per sample. This method reduces complexity, improving pattern learning and generalization, and has shown particular benefit in ECG classification by enhancing the detection of QRS complexes.

Although the raw method provides richer temporal detail, the averaging approach offers lower computational complexity and more stable feature representation, especially beneficial for tasks with relatively stationary or periodic signal features.

### *B. Classifier Models*

We used CNN, LSTM, RSNN, and aRSNN for both tasks; the ECG pipeline also included an MLP with temporal averaging. We compared classification accuracy, computational cost, and training time. This setup enabled us to evaluate performance in terms of accuracy, computational efficiency, and training complexity.

**Non-Spiking Classifiers:**

CNN: The architecture starts with a Conv2D layer (32 filters, 11×11 kernel, ReLU activation), followed by a 3×3 MaxPooling2D and BatchNormalization layer. This pattern repeats with two additional Conv2D layers (3×3 kernel), followed by Flatten, a 128-unit fully connected layer with ReLU activation, and a SoftMax output layer for classification.

LSTM: This model includes an input layer, an LSTM layer with 128 units, followed by global average pooling to reduce dimensionality. A 128-unit fully connected layer with ReLU activation is followed by dropout (rate = 0.5) and a SoftMax output layer for classification.

MLP: The MLP model is combined with the Temporal Averaging Method to provide a lightweight, power-efficient solution. The MLP architecture includes an input layer for the 120-dimensional vector, followed by a hidden layer with 256 units (ReLU), a second hidden layer with 64 units, and a SoftMax output layer.

**Spiking Classifiers:**

RSNN: Comprising two layers with 128 Leaky Integrate-and-Fire (LIF) neurons, this network has event-driven recurrent connections and is trained with Backpropagation Through Time (BPTT) using surrogate gradient descent, optimized by the Adamax algorithm [20].

aRSNN: Built similarly to the RSNN but with Adaptive Leaky Integrate-and-Fire (aLIF) neurons, this model leverages neuron adaptivity for enhanced performance. It uses categorical cross-entropy loss, with Adamax as the optimizer [20].

### *C. Model Training*

**Non-optimized Models (LSTM, CNN, RSNN, aRSNN):** Models were trained with a batch size of 128 over 100 epochs, using a learning rate of 1e-2 for the LSTM, CNN, and RSNN, and 5e-4 for the aRSNN.

**Temporal-averaging MLP:** The MLP was trained with a batch size of 64, a learning rate of 1e-3, and a 10% reduction every 50 epochs over 400 epochs.

### *D. Datasets*

ECG Classification: We used the 2017 PhysioNet/CinC Challenge dataset, containing single-lead ECG recordings categorized as Normal (N), Atrial fibrillation (A), Other rhythms (O), and Noisy signals (~). Data augmentation balanced the dataset by replicating A and N signals, resulting in 1,000 samples per class for training and 250 samples per class for testing.

Sound Classification: The Google Speech Commands (GSC2) dataset provided 6,000 samples of spoken digits for training and testing (5:1 ratio).

### *E. Performance Metrics*

To comprehensively evaluate the classification performance of both spiking and non-spiking models, we employed standard metrics, including precision, recall, and F1 score, alongside accuracy. These metrics provide deeper insights into the models' predictive reliability, especially in imbalanced datasets.

Precision (P): quantifies the proportion of correctly classified positive instances among all predicted positives, assessing the model's ability to minimize false positives:

$$P = \frac{TP}{TP+FP} \tag{2}$$

where $TP$ represents true positives and $FP$ denotes false positives.

Recall (R): measures the proportion of correctly identified positive instances among all actual positives, reflecting the model's ability to detect relevant patterns:

$$R = \frac{TP}{TP+FN} \tag{3}$$

where $FN$ represents false negatives.

F1 Score provides a harmonic mean of precision and recall, balancing both metrics to account for trade-offs between false positives and false negatives:

$$F1 = 2\,\frac{P \times R}{P+R} \tag{4}$$

The metrics in (2)–(4) provide a rigorous assessment of classifier performance across different spike rates and classification tasks. A high F1 score indicates a well-balanced model, particularly in cases where class distributions are uneven. The results, including accuracy and training efficiency, are reported in Section V.

## V. Experimental Results of Real-Life Applications

This section evaluates the performance of the classifiers across ECG and sound classification tasks, examining trade-offs between spiking and non-spiking models and demonstrating Spiketrum's versatility in practical applications.

### *A. ECG Classification*

#### *1) Classification Results*

Among the non-spiking models, the LSTM achieved the highest accuracy, consistently exceeding 80% across most spike rates (Fig. 6(a)). It reached peak accuracy of approximately 85% at 256 spikes per segment, requiring fewer than 60 epochs for convergence. However, the LSTM exhibited higher variability in training performance across different spike rates, suggesting sensitivity to spike distribution patterns. The CNN, while competitive, performed slightly below the LSTM, achieving around 81% accuracy at 128 sps.

Unlike LSTM, CNN required more training epochs to achieve effective convergence, especially at lower spike rates. These differing training requirements highlight the trade-off between the CNN's learning efficiency and its ability to generalize across varied spike rates. For spiking models, the RSNN demonstrated fluctuating performance, achieving a maximum accuracy of approximately 53% at 256 sps. However, it required around 54 epochs for convergence, indicating slower training dynamics compared to the non-spiking models (Fig. 6(b)). In contrast, the aRSNN, which incorporates adaptive dynamics, displayed superior performance among spiking models, reaching nearly 65% accuracy at 512 sps. The aRSNN also exhibited more stable training behavior, with accuracy consistently increasing as the spike rate increased.

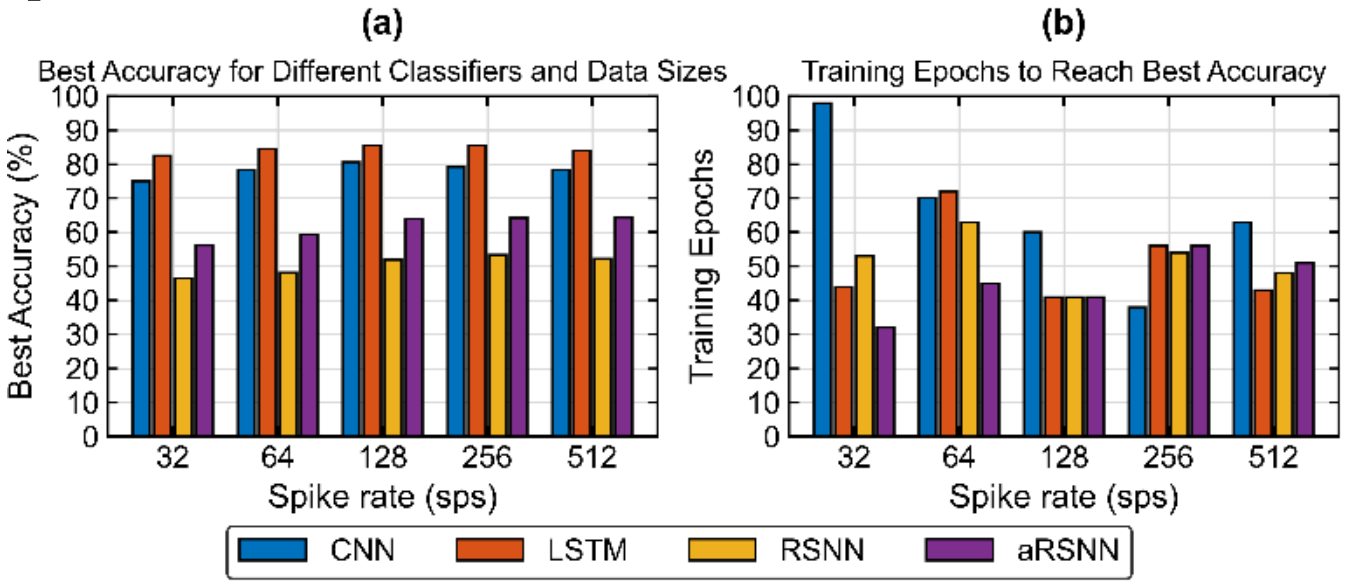

**Fig. 6.** ECG classification accuracy and training epochs across Spiketrum spike rates (spikes per segment; 43.5-ms segments) for spiking and non-spiking classifiers. **(a)** illustrates the maximum classification accuracy on the test dataset after 100 epochs of training for CNN, LSTM, RSNN, and aRSNN. **(b)** illustrates the number of epochs required to achieve the maximum accuracy across various spike rates. These results are based on the analysis performed using the 2017 PhysioNet/CinC Challenge dataset.

The aRSNN required fewer epochs than the RSNN only at 32 and 64 sps; at higher rates, their training times were similar. Adaptive dynamics improved aRSNN accuracy over the RSNN, although both spiking models remained below the non-spiking models. These results highlight the differing sensitivities of spiking and non-spiking models to spike rate configurations. Non-spiking models, particularly LSTM, leverage their sequential processing capabilities to achieve high accuracy but require careful training parameter tuning to avoid variability. Spiking models, while generally less accurate, demonstrate promise for energy-efficient applications, with the aRSNN emerging as the most robust and adaptable classifier due to its ability to optimize learning across varying spike rates with stable and efficient training dynamics.

### *2) Performance Evaluation*

During testing, each model's performance was evaluated on the test set at the optimal spike rate identified during training. Precision, recall, and F1 score were used to assess generalization, with results summarized in Table III.

The table highlights significant differences in classification performance among the evaluated models. The LSTM classifier achieves the highest F1 score (85.30%), with an 8.01-percentage-point advantage over CNN, demonstrating strong predictive accuracy with well-balanced precision and recall. The CNN scored 77.29% F1, with precision and recall of 78.56% and 78.50%, respectively.

TABLE III
PRECISION, RECALL, AND F1 SCORE COMPARISON OF SPIKING AND NON-SPIKING CLASSIFIERS FOR ECG CLASSIFICATION

| Classifiers | Precision (%) | Recall (%) | F1 Score (%) |
|---|---|---|---|
| LSTM | 85.29 | 85.63 | 85.30 |
| CNN | 78.56 | 78.5 | 77.29 |
| RSNN | 54.30 | 53.0 | 53.64 |
| aRSNN | 64.10 | 63.8 | 63.95 |

Among the spiking models, the aRSNN outperforms the RSNN across all metrics, achieving an F1 score of 63.95%, a 10.31-percentage-point improvement over the RSNN's 53.64%. This improvement highlights the benefit of adaptive dynamics in enhancing classification accuracy. However, both spiking models show significantly lower scores compared to non-spiking models, with the RSNN exhibiting the weakest performance, particularly in recall (53.0%), suggesting a tendency to miss positive instances more frequently.

Overall, the results emphasize the superior classification capability of non-spiking models, particularly the LSTM, while demonstrating the potential of adaptive spiking networks in narrowing the performance gap. The lower F1 scores of RSNN and aRSNN indicate that further optimizations, such as architectural enhancements or improved training strategies, may be necessary to enhance their predictive reliability.

### *3) ECG Optimization*

The temporal averaging method, combined with an MLP, was introduced as a lightweight, power-efficient approach for ECG classification, as described in Section IV.

We compare the effectiveness of this optimization by comparing it with conventional deep learning and spiking models using the raw spike input method. The evaluation considers computational complexity, accuracy, and F1 score.

As shown by Table IV, the temporal averaging method, combined with an MLP classifier, achieves the highest accuracy of 93.27% and the best F1 score of 93.75%, while requiring only 95,364 multiply-accumulate operations (MACs). This demonstrates that temporal averaging significantly reduces computational demands while enhancing classification performance, making it a highly efficient approach for resource-constrained environments.

The raw spike input method significantly increases computational complexity across classifiers. While the LSTM and CNN models achieve competitive accuracy (85.50% and 80.62%, respectively), they require 4.8 million and 6.2 million MACs, making them computationally expensive compared to the MLP. The spiking classifiers, RSNN and aRSNN, demand even higher complexity (9.93 million and 11.59 million MACs, respectively) but achieve the lowest accuracy, with the aRSNN gaining 12 percentage points in accuracy over the RSNN. Small differences between Tables III and IV (for example, LSTM F1 scores of 85.42% and 85.30%) arise because Table III reports each model at its best spike rate, whereas Table IV uses a common operating point.

Overall, the results indicate that the combination of temporal averaging and MLP provides the best balance between computational efficiency and classification accuracy. While deep learning models like LSTM and CNN perform well, their higher computational cost makes them less efficient. The spiking networks, although promising for neuromorphic applications, require further optimization to close the accuracy gap relative to the non-spiking classifiers.

TABLE IV
ECG CLASSIFICATION PERFORMANCE AND MODEL COMPLEXITY

| Preprocessing Method | Classifier (256 sps) | Computational Complexity (MACs) | Accuracy (%) | F1 Score (%) |
|---|---|---|---|---|
| Temporal Averaging | MLP | 95,364 | 93.27 | 93.75 |
| Raw Spike Input | LSTM | 4,797,696 | 85.50 | 85.42 |
| | CNN | 6,226,944 | 80.62 | 77.29 |
| | RSNN | 9,932,800 | 51.95 | 51.40 |
| | aRSNN | 11,593,670 | 63.95 | 63.40 |

*B. Sound Classification*

*1) Classification Results*

CNNs achieve consistently high accuracy across all spike rates, exceeding 80% in most cases, with peak accuracy observed at 512 sps (Fig. 7(a)). However, they require a greater number of training epochs. Training required 53 epochs at 32 sps and nearly 100 at 256 sps. Higher spike rates tend to increase CNN training time.

LSTMs maintain stable accuracy across all spike rates, surpassing 85% at 256, 512, and 1024 sps. They require a moderate number of training epochs, with convergence times peaking at 73 epochs at 256 sps. Unlike CNNs, LSTMs demonstrate less sensitivity to spike rate variations, making them more adaptable across different input configurations. Their ability to capture long-term dependencies in sequential data contributes to their robust performance in sound classification tasks (Fig. 7(a)).

Spiking models exhibit greater sensitivity to spike rate variations, highlighting the challenges of spike-based learning. RSNN accuracy peaked at 71% at 512 and 1024 sps but fell below 62% at 32 sps. While it converges relatively quickly at 256 sps, requiring around 45 epochs, it demands significantly more training time at other rates, reflecting inefficiencies in adapting to varying spike densities.

The aRSNN improves upon the RSNN, with accuracy of at least 64.6% at every spike rate. However, it requires more training epochs, particularly at 32 and 64 sps, where its convergence time matches or exceeds that of the RSNN (Fig. 7(b)). Despite its adaptive dynamics offering better stability and accuracy than the RSNN, the aRSNN still struggles to generalize effectively at lower spike rates, where spike-based learning is inherently more challenging.

At lower spike rates (32 and 64 sps), all classifiers, particularly spiking models, exhibit suboptimal performance, indicating a need for careful tuning or alternative preprocessing strategies to enhance learning in low-density spike scenarios. Mid-range spike rates (128 and 256 sps) provide a better balance between accuracy and training efficiency, particularly for LSTMs and CNNs, which leverage these rates to achieve stable performance with manageable training times.

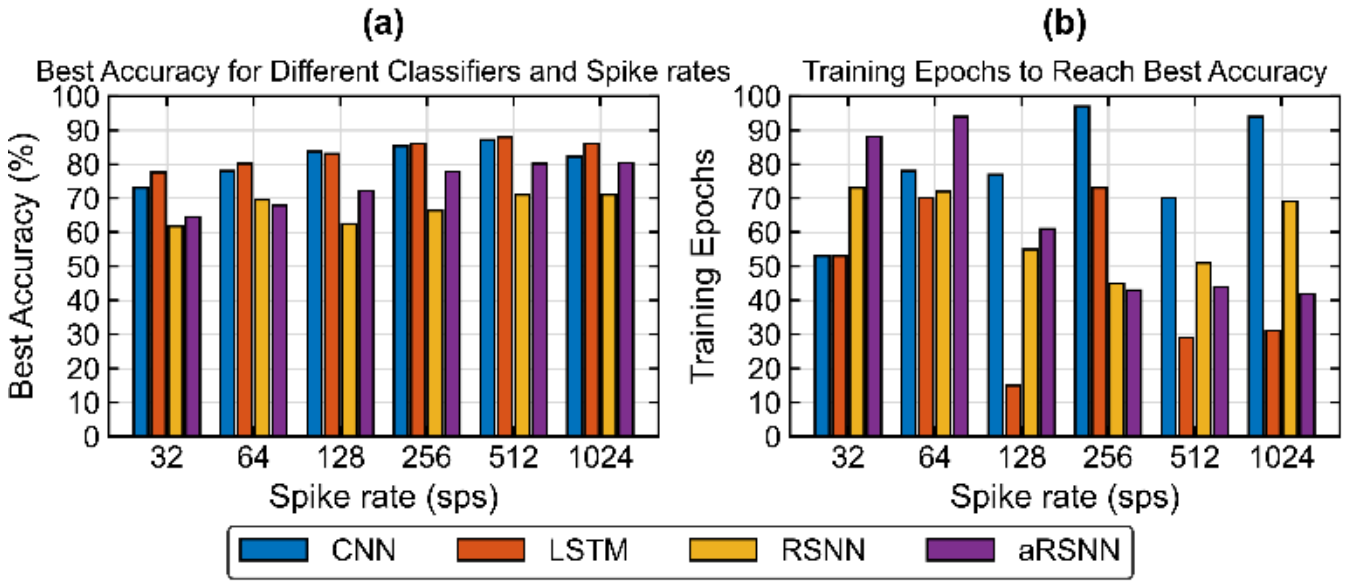

**Fig. 7.** Sound classification accuracy and training time for spiking and non-spiking classifiers across spike rates. **(a)** Best accuracy. **(b)** Epochs required to reach it.

Overall, LSTMs emerge as the most versatile classifier, offering a reliable balance of accuracy and training efficiency across all spike rates. Their ability to process sequential data and capture long-term dependencies makes them particularly suited for sound classification tasks. In contrast, RSNNs and aRSNNs, while capable of faster convergence at specific spike rates, require careful tuning to mitigate performance weaknesses at low spike rates. These results underscore the importance of selecting an appropriate classifier based on the computational constraints and accuracy requirements of the target application, with non-spiking models being more robust and adaptable to varied spike configurations, while spiking models may find utility in energy-constrained scenarios with further optimization.

TABLE V
PERFORMANCE METRICS OF SPIKING AND NON-SPIKING CLASSIFIERS FOR SOUND CLASSIFICATION AT DIFFERENT SPIKE RATES

| Spike rate | Classifier | Precision (%) | Recall (%) | F1 Score (%) |
|---|---|---|---|---|
| 128 | CNN | 81.90 | 80.4 | 80.36 |
| | LSTM | 79.16 | 78.6 | 78.50 |
| | RSNN | 59.09 | 58.9 | 58.84 |
| | aRSNN | 72.10 | 72.0 | 71.88 |
| 256 | CNN | 84.52 | 84.3 | 84.28 |
| | LSTM | 84.92 | 83.8 | 83.76 |
| | RSNN | 63.40 | 62.6 | 63.0 |
| | aRSNN | 76.54 | 76 | 76.01 |

*2) Performance Evaluation*

Here, we compare the performance of non-spiking (CNN and LSTM) and spiking (RSNN and aRSNN) classifiers for sound classification at different spike rates. Table V reports precision, recall, and F1 scores at 128 and 256 sps.

At 128 sps, the CNN achieves the highest F1 score (80.36%), followed closely by the LSTM at 78.50%. Both non-spiking models demonstrate strong performance in precision and recall, with the CNN slightly leading in precision (81.90%). In contrast, spiking models perform less effectively, with the aRSNN achieving an F1 score of 71.88% and the RSNN lagging behind at 58.84%, highlighting the limitations of spiking models at lower spike rates despite some improvements in adaptivity with the aRSNN.

At 256 sps, all classifiers see performance gains. The LSTM had the highest precision (84.92%), while the CNN had the highest F1 score (84.28% versus 83.76% for the LSTM). The aRSNN also improves significantly at the higher spike rate, reaching an F1 score of 76.01% compared to 71.88% at 128 sps, demonstrating its sensitivity to spike rate configurations. However, the RSNN remains the weakest performer, with an F1 score of 63.0%.

These results indicate that non-spiking models, particularly the LSTM and CNN, consistently outperform spiking models in both precision and recall. Still, the aRSNN shows promise for improvement at higher spike rates, reducing the gap between spiking and non-spiking models.

## VI. Discussion

This study presents an area-optimized implementation of a general-purpose spike encoder using the Spiketrum algorithm. Building on our previous work, which introduced a high-throughput, performance-optimized version for parallel-processing architectures such as FPGAs [13], this section expands on Spiketrum's practical utility by comparing both implementations across dimensions like spike generation time, rate, power consumption, resource usage, and practical applicability. This comparative analysis provides insights to support informed decisions for diverse application needs, with benchmarking conducted on an FPGA for consistency. Notably, these results can guide users in selecting the optimal hardware—whether an FPGA, an IC, or another platform—for their specific requirements.

The time-domain area-optimized design generates 1.8 spikes per 43.5-ms segment (42 spikes/s) with eight lanes and 9.2 spikes per segment (about 212 spikes/s) with 40 lanes. In the eight-lane configuration, it uses 11 DSP slices, 2,551 LUTs, and 123 kB of RAM at an estimated 401 mW. This design trades throughput for a smaller resource footprint and suits low-rate workloads; rates above its 40-lane limit require the frequency-domain implementation for real-time processing.

In contrast, the performance-optimized version, leveraging frequency-domain convolution, achieves a notably fast spike generation time of 0.5 ms, generating up to 80 spikes per segment (1,840 spikes/s). This implementation's performance comes at the cost of significantly higher resource usage—235 DSP slices (2,036% more than the eight-lane area-optimized design), 285 BRAMs (751% more than the eight-lane area-optimized design), and 8,756 LUTs (243% more than the eight-lane area-optimized design)—positioning it for tasks that demand substantial computational power and high-speed processing. At an estimated 1.38 W, with 89% dynamic power, this design favors high-throughput real-time tasks over energy-constrained use.

Table VI compares Spiketrum's FPGA implementations with other spike encoders from the literature, emphasizing differences in power efficiency, resource utilization, and throughput.

TABLE VI
A Comparison of Spiketrum Implementations with Other Encoders Found in the Literature

| Design | Platform | Power (mW) | DSP | LUTs | RAM (kB) |
|---|---|---|---|---|---|
| Area-optimized Spiketrum, L=8 | Artix-7 | 401 | 11 | 2,551 | 123 |
| IC, L=1 | 180 nm | 63 | – | – | – |
| Performance-optimized Spiketrum [13] | Artix-7 | 1,382 | 235 | 8,756 | 719 |
| Audio [21] | Cyclone | 1,260 | 49 | 5,235 | 571 |
| ECG [22] | Artix-7 | 28 | 21 | 19,663 | 50 |
| EMG [23] | Kintex | – | 498 | 90,683 | 2,894 |

Spiketrum FPGA power values are Vivado estimates; values for other designs are reported in their sources. L denotes MACC lanes; RAM denotes stored data.

The audio-processing encoder in [21] consumes 1,260 mW, which is 8.8% below Spiketrum's 1,382 mW performance-optimized implementation but significantly higher than the area-optimized version (401 mW, 68.2% lower). Despite its lower power consumption compared to the performance-optimized Spiketrum, its throughput remains unspecified, making direct efficiency comparisons challenging. Additionally, this encoder uses 49 DSP slices, which is 79.15% lower than Spiketrum's 235 DSPs, but 345% more than Spiketrum's area-optimized design (11 DSP slices). In terms of LUT utilization, it consumes 5,235 LUTs, 40.2% fewer than Spiketrum's performance-optimized version but 105% more than Spiketrum's area-optimized design. With a RAM usage of 571 kB, this is 20.6% below the performance-optimized version (719 kB) but 364% above the area-optimized version (123 kB).

The ECG-optimized encoder in [22] is highly power-efficient, consuming just 28 mW. Its throughput is reported in beats/s, whereas Spiketrum's is measured in spikes/s, so the rates are not directly comparable. Single-lane IC power is reported separately in Table II.

The EMG-based encoder in [23] is highly resource-intensive, consuming 498 DSP slices and 90,683 LUTs, yet achieving only 2 samples per second. Using reported figures, the performance-optimized Spiketrum uses 52.81% fewer DSP slices and 90.34% fewer LUTs than the EMG encoder while generating up to 1,840 spikes/s. RAM usage is 2,894 kB, which is 302% higher than Spiketrum's performance-optimized version, reflecting the EMG encoder's reliance on extensive memory resources. The area-optimized design uses 11 DSP slices, 2,551 LUTs, and 123 kB of RAM; compared with the EMG encoder, these are 97.8%, 97.2%, and 95.8% lower, respectively.

Compared to alternative designs, Spiketrum provides configurable spike rates; its area-optimized design reduces hardware use for sensory applications with modest throughput requirements. The area-optimized implementation is particularly suited for applications requiring a balance between low resource consumption and low-to-moderate rates, including low-power neural networks and sensor processing, where computational and power efficiency are critical.

The choice between these implementations depends on the specific computational demands and resource constraints of the target application. For tasks requiring minimal power and area, the Spiketrum IC implementation is the most suitable option. It achieves an optimal balance between energy efficiency and compact design, making it ideal for energy-sensitive environments and applications with strict area constraints. Meanwhile, the performance-optimized FPGA implementation is better suited for high-performance scenarios, where speed and precision outweigh power and area limitations.

While the IC implementation incorporates a variety of optimization techniques, certain limitations persist. The use of the 180 nm process technology, although robust and well-established, lacks the power efficiency and higher transistor density available in more advanced nodes, such as 7 nm or 5 nm technologies. Transitioning to these smaller process nodes could further improve power efficiency and compactness. Additionally, off-chip storage for Signal RAM and Time-Domain Kernel ROM effectively reduces on-chip area requirements but introduces potential latency and bandwidth bottlenecks, particularly in high-throughput scenarios. Furthermore, the inclusion of numerous I/O pins, driven by testing and verification requirements, limits layout efficiency and increases the chip's overall footprint.

Despite these challenges, the selected technology and design approach strike a practical balance between feasibility, reliability, and cost-effectiveness. The Spiketrum IC implementation, with its compact die size and low power consumption, is particularly well-suited for prototyping and low-to-medium volume production, while maintaining the adaptability to address a variety of neuromorphic and signal processing challenges across diverse application domains.

## VII. Conclusion

This study presented a scalable, time-domain Spiketrum design in which lane count trades throughput for hardware cost while preserving identical output codes. Compared with the frequency-domain performance-optimized design, the eight-lane configuration uses 88% fewer BRAMs, 95% fewer DSP slices, and 71% fewer LUTs, with 71% lower Vivado-estimated power. In application tests, an MLP with temporal averaging achieved 93.27% ECG accuracy using 95,364 MACs, while an LSTM achieved 87.9% sound-classification accuracy at 512 spikes per segment. Physical design of the single-lane core in TSMC 180-nm technology provides an IC proof of concept.

Spiketrum could serve as a configurable sensory front end, encoding high-volume signals near their source into compact spike representations for spiking or conventional downstream models. By matching the spike rate to the task and available hardware, it could reduce data movement and processing demands in edge AI, wearable health monitoring, robotics, autonomous systems, and future multimodal sensing. Future work will validate this potential across larger real-world datasets and modalities and evaluate the IC through system-level and hardware–software co-design studies.

## References

[1] G. S. Rose, M. S. A. Shawkat, A. Z. Foshie, J. J. Murray, and M. M. Adnan, "A system design perspective on neuromorphic computer processors," *Neuromorphic Computing and Engineering,* vol. 1, no. 2, p. 022001, 2021.

[2] C. D. Schuman, J. S. Plank, G. Bruer, and J. Anantharaj, "Non-traditional input encoding schemes for spiking neuromorphic systems," in *2019 International Joint Conference on Neural Networks (IJCNN)*, 2019: IEEE, pp. 1-10.

[3] X. Qi, X. Guo, and J. G. Harris, "A time-to-first spike CMOS imager," in *2004 IEEE International Symposium on Circuits and Systems (ISCAS)*, 2004, vol. 4: IEEE, pp. IV-824.

[4] C. Shoushun and A. Bermak, "Arbitrated time-to-first spike CMOS image sensor with on-chip histogram equalization," *IEEE Transactions on Very Large Scale Integration (VLSI) Systems,* vol. 15, no. 3, pp. 346-357, 2007.

[5] Y. Yi *et al.*, "FPGA based spike-time dependent encoder and reservoir design in neuromorphic computing processors," *Microprocessors and Microsystems,* vol. 46, pp. 175-183, 2016.

[6] S. M. Bohte, J. N. Kok, and H. La Poutre, "Error-backpropagation in temporally encoded networks of spiking neurons," *Neurocomputing,* vol. 48, no. 1-4, pp. 17-37, 2002.

[7] H. Mostafa, "Supervised learning based on temporal coding in spiking neural networks," *IEEE transactions on neural networks and learning systems,* vol. 29, no. 7, pp. 3227-3235, 2017.

[8] C. Liu *et al.*, "A spiking neuromorphic design with resistive crossbar," in *Proceedings of the 52nd Annual Design Automation Conference*, 2015, pp. 1-6.

[9] A. Bagheri, O. Simeone, and B. Rajendran, "Training probabilistic spiking neural networks with first-to-spike decoding," in *2018 IEEE International Conference on Acoustics, Speech and Signal Processing (ICASSP)*, 2018: IEEE, pp. 2986-2990.

[10] D. Querlioz, O. Bichler, and C. Gamrat, "Simulation of a memristor-based spiking neural network immune to device variations," in *The 2011 International Joint Conference on Neural Networks*, 2011: IEEE, pp. 1775-1781.

[11] C. Schuman, C. Rizzo, J. McDonald-Carmack, N. Skuda, and J. Plank, "Evaluating encoding and decoding approaches for spiking neuromorphic systems," in *Proceedings of the International Conference on Neuromorphic Systems 2022*, 2022, pp. 1-9.

[12] H. Tang *et al.*, "Neuromorphic auditory perception by neural spiketrum," *IEEE Transactions on Emerging Topics in Computational Intelligence,* 2024.

[13] M. A. Alsakkal and J. Wijekoon, "Spiketrum: An FPGA-Based Implementation of a Neuromorphic Cochlea," *IEEE Transactions on Circuits and Systems,* 2025, doi: 10.1109/TCSI.2025.3526585.

[14] S. G. Mallat and Z. J. I. T. o. s. p. Zhang, "Matching pursuits with time-frequency dictionaries," vol. 41, no. 12, pp. 3397-3415, 1993.

[15] R. M. Bittner, J. Salamon, M. Tierney, M. Mauch, C. Cannam, and J. P. Bello, "Medleydb: A multitrack dataset for annotation-intensive mir research," in *ISMIR*, 2014, vol. 14, pp. 155-160.

[16] M. A. Alsakkal, "Application based Evaluation of an Efficient Spike-Encoder, "Spiketrum"," *Under review with IEEE Transactions on Emerging Topics in Computational Intelligence,* 2026.

[17] A. Finnerty and H. Ratigner, "Reduce power and cost by converting from floating point to fixed point," *WP491 (v1. 0),* 2017.

[18] Xilinx, "7 Series DSP48E1 Slice User Guide," 2018. [Online]. Available: https://docs.xilinx.com/v/u/en-US/ug479_7Series_DSP48E1

[19] T. Feist, "Vivado design suite," *White Paper,* vol. 5, p. 30, 2012.

[20] D. P. Kingma and J. Ba, "Adam: A method for stochastic optimization," *arXiv preprint arXiv:1412.6980,* 2014.

[21] Y. Xu, C. S. Thakur, R. K. Singh, T. J. Hamilton, R. M. Wang, and A. Van Schaik, "A FPGA implementation of the CAR-FAC cochlear model," *Frontiers in neuroscience,* vol. 12, p. 198, 2018.

[22] M. P. Desai, G. Caffarena, R. Jevtic, D. G. Márquez, and A. Otero, "A low-latency, low-power FPGA implementation of ECG signal characterization using hermite polynomials," *Electronics,* vol. 10, no. 19, p. 2324, 2021.

[23] H.-S. Choi, "Electromyogram (EMG) signal classification based on light-weight neural network with FPGAs for wearable application," *Electronics,* vol. 12, no. 6, p. 1398, 2023.